\documentclass[12pt]{article}%
\usepackage{amsfonts}
\usepackage{sw20bams}
\usepackage{amsmath}
\usepackage{amssymb}
\usepackage{graphicx}%
\providecommand{\U}[1]{\protect\rule{.1in}{.1in}}
\begin{document}

\title{Resolving Conflicts and Electing Leaders}
\author{Steven Finch}
\date{December 9, 2019}
\maketitle

\begin{abstract}
We review distributed algorithms for transmitting data ($n$ real numbers)
under a broadcast communication model, as well as for maximum finding and for
sorting. \ Our interest is in the basics of recursive formulas and
corresponding asymptotics as $n\rightarrow\infty$. \ The emphasis is on
concrete examples rather than general theory.

\end{abstract}

\footnotetext{Copyright \copyright \ 2019 by Steven R. Finch. All rights
reserved.}Assume that $n$ persons communicate via a single channel and each
individual possesses a real number that he/she urgently wishes to share with
everyone else. \ Unfortunately the channel can transmit only one item at a
time; although communication is instantaneous, it is limited by the discrete
nature of time. \ Everyone attempts to broadcast at the beginning, of course.
\ A \textbf{conflict} arises. \ No transmission can occur until the conflict
is \textbf{resolved}. \ The persons involved in any conflict (large or small)
toss independent random fair coins; those who obtain heads step out of the
game and wait until those who obtained tails proceed with a new round of coin
tosses. \ If no tails were obtained, then the old round must be repeated.
\ Eventually a unique person emerges, the \textbf{leader} of an
\textbf{election}, and is allowed to transmit his/her number. \ There are
still $n-1$ individuals left at this point. \ Starting afresh is a possible
next step, but this would be needlessly inefficient. \ The strongly binary
tree already constructed in the preceding rounds (a natural history of
consecutive coin tosses) can be efficiently further grown, permitting all
$n-1$ transmissions to ultimately occur one-by-one \cite{Ms-tcs1, Ya-tcs1,
MP-tcs1}. \ Quantifying the final number of vertices in the tree is important,
along with the length of the root-to-leaf path that initially gave the leader. \ 

We discuss details in Sections 1, 2 and 3. \ Relevant distributed or parallel
algorithms for finding the maximum of the $n$ numbers and for sorting the $n$
numbers are covered in Sections 4 and 5. \ Note throughout that knowledge of
the integer $n$ is not required for the algorithms to execute; while the
analyst is aware of the value of $n$, the $n$ individuals do not.

\section{Conflict Resolutions}

Consider the R\ program:%
\[%
\begin{array}
[c]{l}%
\text{\texttt{f
$<$%
- function(L,k)}}\\
\text{\texttt{\{}}\\
\text{\texttt{\ \ if(NROW(L)%
$<$%
=1) \{}}\\
\text{\texttt{\ \ \ \ k}}\\
\text{\texttt{\ \ \} else \{}}\\
\text{\texttt{\ \ \ \ b
$<$%
- rbinom(NROW(L),1,0.5)}}\\
\text{\texttt{\ \ \ \ L1
$<$%
- L[(1:NROW(L))*(1-b)]}}\\
\text{\texttt{\ \ \ \ k
$<$%
- f(L1,k
$<$%
- k+1)}}\\
\text{\texttt{\ \ \ \ L2
$<$%
- L[(1:NROW(L))*b]}}\\
\text{\texttt{\ \ \ \ k
$<$%
- f(L2,k
$<$%
- k+1)}}\\
\text{\texttt{\ \ \}}}\\
\text{\texttt{\ \ k}}\\
\text{\texttt{\}}}%
\end{array}
\]
where the integer $k$ is initially $1$, the list $L$ is initially
$\{1,2,\ldots,n\}$ for simplicity's sake, and $\operatorname{rbinom}%
(\cdot,1,0.5)$ is the built-in R unbiased random Bernoulli $\{0,1\}$
generator. \ Figure 1 exhibits a sample binary tree for $n=5$. \ The first
round of coin tosses yields $b=(0,1,0,1,1)$, where $0=\operatorname{tail}$ and
$1=\operatorname{head}$, and thus $L_{1}=(1,3)$, $L_{2}=(2,4,5)$. \ On the
left side, focusing on $L_{1}$ only, the next round of tosses yields
$b^{\prime}=(1,1)$ and thus $L_{1}^{\prime}=\emptyset$, $L_{2}^{\prime
}=\{1,3\}$. \ On the right side, focusing on $L_{2}$ only, the next round of
tosses yields $b^{\prime\prime}=(0,0,0)$ and thus $L_{1}^{\prime\prime
}=\{2,4,5\}$, $L_{2}^{\prime\prime}=\emptyset$. \ Also, focusing on
$L_{1}^{\prime\prime}$ only, the next round of tosses yields $b^{\prime
\prime\prime}=(0,0,1)$ and thus $L_{1}^{\prime\prime\prime}=\{2,4\}$,
$L_{2}^{\prime\prime\prime}=\{5\}$. \ Termination occurs at the next level:
any twig of the tree ceases to grow when its leaf contains $\leq1$ items.
\ The number of vertices (including empty vertices) is $13$. \ We write
$X_{5}=13$. \ The labeled ordering of vertices is determined by watching $k$
increase, from $1$ to $13$, as the algorithm progresses. \ Smaller values of
$k$ occur on the left side because recursive splitting starts with $L_{1}$,
$L_{2}^{\prime}$, \ldots\ and thereafter continues with $L_{2}$,
$L_{1}^{\prime\prime}$, $L_{1}^{\prime\prime\prime}$, \ldots.%
\begin{figure}[ptb]%
\centering
\includegraphics[
height=6.337in,
width=6.1461in
]%
{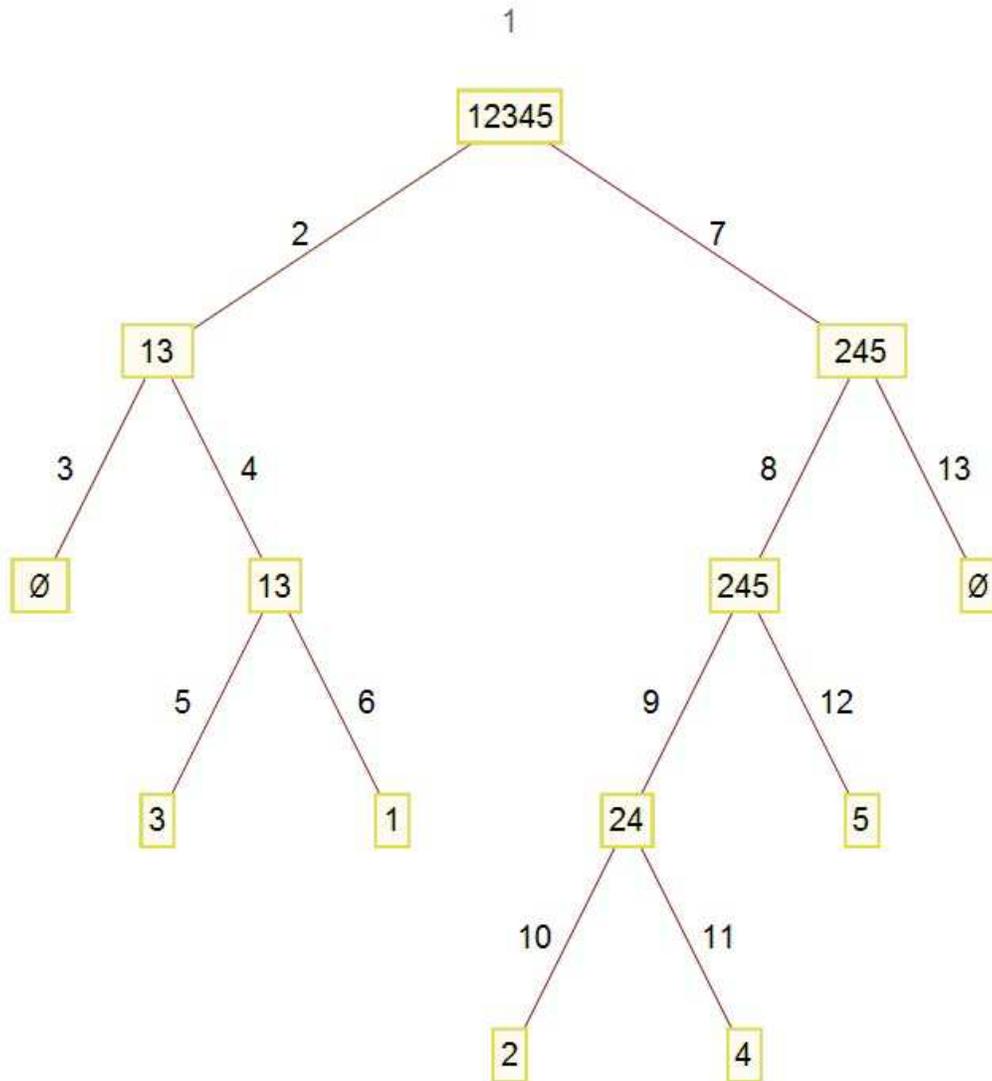}%
\caption{Strongly binary tree for conflict resolution.}%
\end{figure}

The probability generating function for $X_{n}$, given $n$, is likewise
recursive \cite{MP-tcs1}:%
\[%
\begin{array}
[c]{ccc}%
f_{n}(z)=z2^{-n}%
{\displaystyle\sum\limits_{k=0}^{n}}
\dbinom{n}{k}f_{k}(z)f_{n-k}(z), &  & n\geq2;
\end{array}
\]%
\[
f_{0}(z)=f_{1}(z)=z.
\]
Note that $f_{n}(1)=1$ always. \ Differentiating with respect to $z$:%
\[
f_{n}^{\prime}(z)=2^{-n}%
{\displaystyle\sum\limits_{k=0}^{n}}
\dbinom{n}{k}f_{k}(z)f_{n-k}(z)+z2^{-n}%
{\displaystyle\sum\limits_{k=0}^{n}}
\dbinom{n}{k}\left[  f_{k}^{\prime}(z)f_{n-k}(z)+f_{k}(z)f_{n-k}^{\prime
}(z)\right]
\]
we have first moment%
\begin{align*}
\mathbb{E}(X_{n})  &  =f_{n}^{\prime}(1)=1+2^{-n}%
{\displaystyle\sum\limits_{k=0}^{n}}
\dbinom{n}{k}\left[  f_{k}^{\prime}(1)+f_{n-k}^{\prime}(1)\right] \\
&  =1+2^{1-n}%
{\displaystyle\sum\limits_{k=0}^{n}}
\dbinom{n}{k}f_{k}^{\prime}(1),
\end{align*}
that is,
\begin{align*}
g_{n}  &  =1+2^{1-n}%
{\displaystyle\sum\limits_{k=0}^{n}}
\dbinom{n}{k}g_{k}\\
&  =\frac{1}{1-2^{1-n}}\left[  1+2^{1-n}%
{\displaystyle\sum\limits_{k=0}^{n-1}}
\dbinom{n}{k}g_{k}\right]
\end{align*}
where $g_{k}=f_{k}^{\prime}(1)$ and $g_{0}=g_{1}=1$. \ Clearly $g_{2}=5$ and
$g_{3}=23/3$. \ Differentiating again:%
\begin{align*}
f_{k}^{\prime\prime}(z)  &  =2^{1-n}%
{\displaystyle\sum\limits_{k=0}^{n}}
\dbinom{n}{k}\left[  f_{k}^{\prime}(z)f_{n-k}(z)+f_{k}(z)f_{n-k}^{\prime
}(z)\right] \\
&  +2^{-n}%
{\displaystyle\sum\limits_{k=0}^{n}}
\dbinom{n}{k}\left[  f_{k}^{\prime\prime}(z)f_{n-k}(z)+2f_{k}^{\prime
}(z)f_{n-k}^{\prime}(z)+f_{k}(z)f_{n-k}^{\prime\prime}(z)\right]
\end{align*}
we have second factorial moment%
\begin{align*}
\mathbb{E}(X_{n}(X_{n}-1))  &  =f_{n}^{\prime\prime}(1)\\
&  =-2+2f_{n}^{\prime}(1)+2^{1-n}%
{\displaystyle\sum\limits_{k=0}^{n}}
\dbinom{n}{k}f_{k}^{\prime}(1)f_{n-k}^{\prime}(1)+2^{1-n}%
{\displaystyle\sum\limits_{k=0}^{n}}
\dbinom{n}{k}f_{k}^{\prime\prime}(1),
\end{align*}
that is,%
\begin{align*}
h_{n}  &  =-2+2g_{n}+2^{1-n}%
{\displaystyle\sum\limits_{k=0}^{n}}
\dbinom{n}{k}g_{k}g_{n-k}+2^{1-n}%
{\displaystyle\sum\limits_{k=0}^{n}}
\dbinom{n}{k}h_{k}\\
&  =\frac{1}{1-2^{1-n}}\left[  -2+2g_{n}+2^{1-n}%
{\displaystyle\sum\limits_{k=0}^{n}}
\dbinom{n}{k}g_{k}g_{n-k}+2^{1-n}%
{\displaystyle\sum\limits_{k=0}^{n-1}}
\dbinom{n}{k}h_{k}\right]
\end{align*}
where $h_{k}=f_{k}^{\prime\prime}(1)$ and $h_{0}=h_{1}=0$. \ Clearly
$h_{2}=28$ and $h_{3}=548/9$. \ Finally, we have variance%
\[
\mathbb{V}(X_{n})=h_{n}-g_{n}^{2}+g_{n}%
\]
which is $8$ when $n=2$ and $88/9$ when $n=3$. \ 

Such recursions are helpful for small $n$, but give no asymptotic information
as $n\rightarrow\infty$. \ It can be proved that \cite{PS-tcs1, FHZ-tcs1}%
\begin{align*}
\frac{\mathbb{E}(X_{n})}{n}  &  =\frac{2}{\ln(2)}+\delta_{1}\left(  \frac
{\ln(n)}{\ln(2)}\right)  +o(1)\\
&  \approx(2.8853900817...)+\,\delta_{1}\left(  \log_{2}(n)\right)  ,
\end{align*}%
\begin{align*}
\frac{\mathbb{V}(X_{n})}{n}  &  =\frac{1}{\ln(2)}\left(  1+8%
{\displaystyle\sum\limits_{k=1}^{\infty}}
\frac{1}{(2^{k}+1)^{2}}\right)  +\delta_{2}\left(  \frac{\ln(n)}{\ln
(2)}\right)  +o(1)\\
&  \approx(3.3834344923...)+\,\delta_{2}\left(  \log_{2}(n)\right)
\end{align*}
where $\delta_{j}(x)$ is a periodic function of period $1$ with tiny amplitude
and zero mean, for $j=1$, $2$. \ Functions like these appear throughout the
analysis of algorithms. \ We shall not specify these here nor later, but
merely indicate their presence when required. \ An infinite series
representation of the mean for arbitrary $n$ is \cite{FJ-tcs1, Ro-tcs1}%
\[
\mathbb{E}(X_{n})=1+2%
{\displaystyle\sum\limits_{m=0}^{\infty}}
2^{m}\left[  1-\left(  1-2^{-m}\right)  ^{n}-n2^{-m}\left(  1-2^{-m}\right)
^{n-1}\right]  .
\]
Also, the constant $0.8458586230...$ (one-quarter of the variance as
$n\rightarrow\infty$) appears in \cite{RJ-tcs1, KP1-tcs1, KP2-tcs1}.

\section{Leader Elections}

Two parameters are examined in this section:\ \textit{height} (length of the
root-to-leaf path that gives the leader) and \textit{size} (number of vertices
in the associated weakly binary tree). \ We further discuss a variation of the
election in which \textit{draws} (involving exactly two competitors)
constitute an additional way to stop the process.

\subsection{Height}

Consider the R\ program:%
\[%
\begin{array}
[c]{l}%
\text{\texttt{f
$<$%
- function(L,k)}}\\
\text{\texttt{\{}}\\
\text{\texttt{\ \ if(NROW(L)%
$<$%
=1) \{}}\\
\text{\texttt{\ \ \ \ k}}\\
\text{\texttt{\ \ \} else \{}}\\
\text{\texttt{\ \ \ \ b
$<$%
- rbinom(NROW(L),1,0.5)}}\\
\text{\texttt{\ \ \ \ L1
$<$%
- L[(1:NROW(L))*(1-b)]}}\\
\text{\texttt{\ \ \ \ k
$<$%
- f(L1,k
$<$%
- k+1)}}\\
\text{\texttt{\ \ \ \ if(NROW(L1)==0) \{}}\\
\text{\texttt{\ \ \ \ \ \ L2
$<$%
- L[(1:NROW(L))*b]}}\\
\text{\texttt{\ \ \ \ \ \ k
$<$%
- f(L2,k)}}\\
\text{\texttt{\ \ \ \ \}}}\\
\text{\texttt{\ \ \}}}\\
\text{\texttt{\ \ k}}\\
\text{\texttt{\}}}%
\end{array}
\]
where the integer $k$ is initially $0$ (unlike before) and the list $L$ is
initially $\{1,2,\ldots,n\}$. \ Figure 2 exhibits a sample binary tree for
$n=5$. \ The first round of coin tosses yields $b=(0,0,0,1,0)$, where
$0=\operatorname{tail}$ and $1=\operatorname{head}$, and thus $L_{1}%
=(1,2,3,5)$, $L_{2}=(4)$. \ Focusing on $L_{1}$ only, the next round of tosses
yields $b^{\prime}=(1,1,1,1)$ and thus $L_{1}^{\prime}=\emptyset$,
$L_{2}^{\prime}=\{1,2,3,5\}$. \ Note that we do not indicate nor count the
empty vertex. Focusing on $L_{2}^{\prime}$ only, the next round of tosses
yields $b^{\prime\prime}=(1,0,0,0)$ and thus $L_{1}^{\prime\prime}=\{2,3,5\}$,
$L_{2}^{\prime\prime}=\{1\}$. \ Focusing on $L_{1}^{\prime\prime}$ only, the
next round of tosses yields $b^{\prime\prime\prime}=(0,0,0)$ and thus
$L_{1}^{\prime\prime\prime}=\{2,3,5\}$, $L_{2}^{\prime\prime\prime}=\emptyset
$. \ Focusing on $L_{1}^{\prime\prime\prime}$ only, the next round of tosses
yields $b^{\prime\prime\prime\prime}=(0,0,1)$ and thus $L_{1}^{\prime
\prime\prime\prime}=\{2,3\}$, $L_{2}^{\prime\prime\prime\prime}=\{5\}$.
\ Termination occurs at the next level: the leftmost twig of the tree contains
exactly $1$ item. \ The height (number of steps separating the vertices
labeled $1$ and $7$) is $6$. We write $H_{5}=6$. \
\begin{figure}[ptb]%
\centering
\includegraphics[
height=6.3387in,
width=3.9834in
]%
{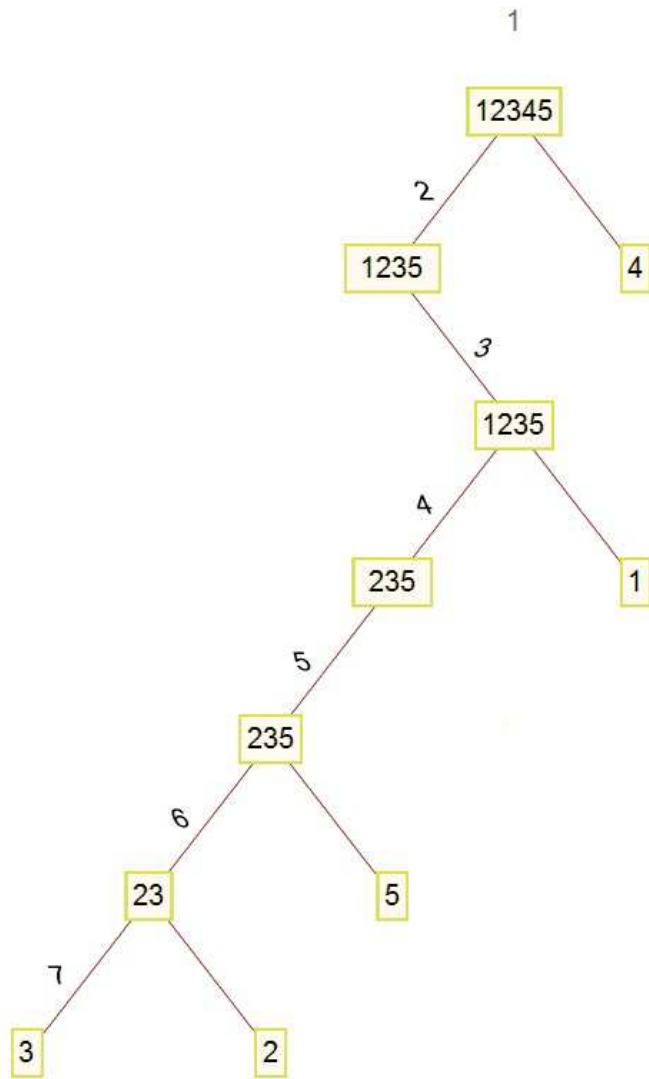}%
\caption{Weakly binary tree for leader election of height 6.}%
\end{figure}

The probability generating function for $H_{n}$, given $n$, is \cite{Pr-tcs1}:%
\[%
\begin{array}
[c]{ccc}%
f_{n}(z)=-z2^{-n}+z2^{-n}f_{n}(z)+z2^{-n}%
{\displaystyle\sum\limits_{k=0}^{n}}
\dbinom{n}{k}f_{k}(z), &  & n\geq2;
\end{array}
\]%
\[
f_{0}(z)=f_{1}(z)=1.
\]
Note that $f_{n}(1)=1$ always. \ Differentiating with respect to $z$:%
\[
f_{n}^{\prime}(z)=-2^{-n}+2^{-n}f_{n}(z)+z2^{-n}f_{n}^{\prime}(z)+2^{-n}%
{\displaystyle\sum\limits_{k=0}^{n}}
\dbinom{n}{k}f_{k}(z)+z2^{-n}%
{\displaystyle\sum\limits_{k=0}^{n}}
\dbinom{n}{k}f_{k}^{\prime}(z)
\]
we have first moment%
\begin{align*}
\mathbb{E}(H_{n})  &  =f_{n}^{\prime}(1)=-2^{-n}+2^{-n}+2^{-n}f_{n}^{\prime
}(1)+1+2^{-n}%
{\displaystyle\sum\limits_{k=0}^{n}}
\dbinom{n}{k}f_{k}^{\prime}(1)\\
&  =1+2^{-n}f_{n}^{\prime}(1)+2^{-n}%
{\displaystyle\sum\limits_{k=0}^{n}}
\dbinom{n}{k}f_{k}^{\prime}(1),
\end{align*}
that is,
\begin{align*}
g_{n}  &  =1+2^{-n}g_{n}+2^{-n}%
{\displaystyle\sum\limits_{k=0}^{n}}
\dbinom{n}{k}g_{k}\\
&  =\frac{1}{1-2^{1-n}}\left[  1+2^{-n}%
{\displaystyle\sum\limits_{k=0}^{n-1}}
\dbinom{n}{k}g_{k}\right]
\end{align*}
where $g_{k}=f_{k}^{\prime}(1)$ and $g_{0}=g_{1}=0$. \ Clearly $g_{2}=2$ and
$g_{3}=7/3$. \ Differentiating again:%
\[
f_{n}^{\prime\prime}(z)=2^{1-n}f_{n}^{\prime}(z)+z2^{-n}f_{n}^{\prime\prime
}(z)+2^{1-n}%
{\displaystyle\sum\limits_{k=0}^{n}}
\dbinom{n}{k}f_{k}^{\prime}(z)+z2^{-n}%
{\displaystyle\sum\limits_{k=0}^{n}}
\dbinom{n}{k}f_{k}^{\prime\prime}(z)
\]
we have second factorial moment%
\begin{align*}
\mathbb{E}(H_{n}(H_{n}-1))  &  =f_{n}^{\prime\prime}(1)\\
&  =-2+2f_{n}^{\prime}(1)+2^{-n}f_{n}^{\prime\prime}(1)+2^{-n}%
{\displaystyle\sum\limits_{k=0}^{n}}
\dbinom{n}{k}f_{k}^{\prime\prime}(1),
\end{align*}
that is,
\begin{align*}
h_{n}  &  =-2+2g_{n}+2^{-n}h_{n}+2^{-n}%
{\displaystyle\sum\limits_{k=0}^{n}}
\dbinom{n}{k}h_{k}\\
&  =\frac{1}{1-2^{1-n}}\left[  -2+2g_{n}+2^{-n}%
{\displaystyle\sum\limits_{k=0}^{n-1}}
\dbinom{n}{k}h_{k}\right]
\end{align*}
where $h_{k}=f_{k}^{\prime\prime}(1)$ and $h_{0}=h_{1}=0$. \ Clearly $h_{2}=4$
and $h_{3}=50/9$. \ Finally, we have variance $\mathbb{V}(H_{n})$ which is $2$
when $n=2$ and $22/9$ when $n=3$. \ 

It can be proved that \cite{FMS-tcs1, JS-tcs1, Ks-tcs1}%
\begin{align*}
\mathbb{E}(H_{n})  &  =\frac{\ln(n)}{\ln(2)}+\frac{1}{2}+\delta_{3}\left(
\frac{\ln(n)}{\ln(2)}\right)  +o(1)\\
&  \approx\log_{2}(n)+(0.5)+\delta_{3}(\log_{2}(n)),
\end{align*}%
\begin{align*}
\mathbb{V}(H_{n})  &  =\frac{1}{12}+\frac{\pi^{2}}{6\ln(2)^{2}}-\frac
{\gamma^{2}+2\gamma_{1}}{\ln(2)^{2}}+\varepsilon\left(  \frac{\ln(n)}{\ln
(2)}\right)  +o(1)\\
&  \approx(3.1166951643...)+\,\varepsilon\left(  \log_{2}(n)\right)
\end{align*}
as $n\rightarrow\infty$, where $\gamma_{1}$ is the first Stieltjes constant
\cite{Fi0-tcs1} and fluctuations provided by $\varepsilon(x)$ are symmetrical
not about $0$, but instead about \cite{Fi2-tcs1}%
\[%
{\displaystyle\int\limits_{0}^{1}}
\varepsilon(x)dx=-\frac{1}{\ln(2)^{2}}%
{\displaystyle\sum\limits_{\substack{k=-\infty\\k\neq0}}^{\infty}}
\left\vert \zeta\left(  1-\frac{2\pi ik}{\ln(2)}\right)  \Gamma\left(
1-\frac{2\pi ik}{\ln(2)}\right)  \right\vert ^{2}\approx-1.856\times10^{-10}.
\]
Needless to say, the evaluation of $\mathbb{V}(H_{n})$ is an impressive and
difficult achievement.

\subsection{Size}

Consider the R\ program:%
\[%
\begin{array}
[c]{l}%
\text{\texttt{f
$<$%
- function(L,k)}}\\
\text{\texttt{\{}}\\
\text{\texttt{\ \ if(NROW(L)%
$<$%
=1) \{}}\\
\text{\texttt{\ \ \ \ k}}\\
\text{\texttt{\ \ \} else \{}}\\
\text{\texttt{\ \ \ \ b
$<$%
- rbinom(NROW(L),1,0.5)}}\\
\text{\texttt{\ \ \ \ L1
$<$%
- L[(1:NROW(L))*(1-b)]}}\\
\text{\texttt{\ \ \ \ L2
$<$%
- L[(1:NROW(L))*b]}}\\
\text{\texttt{\ \ \ \ k
$<$%
- f(L1,k
$<$%
- k+(NROW(L1)!=0)+(NROW(L2)!=0))}}\\
\text{\texttt{\ \ \ \ if(NROW(L1)==0) \{}}\\
\text{\texttt{\ \ \ \ \ \ k
$<$%
- f(L2,k)}}\\
\text{\texttt{\ \ \ \ \}}}\\
\text{\texttt{\ \ \}}}\\
\text{\texttt{\ \ k}}\\
\text{\texttt{\}}}%
\end{array}
\]
where the integer $k$ is initially $1$ and the list $L$ is initially
$\{1,2,\ldots,n\}$. \ Figure 3 is almost identical to Figure 2 -- it contains
the same binary tree for $n=5$ -- but the labeling is different. \ The number
of vertices (excluding empty vertices) is $11$. \ We write $Y_{5}=11$. \ The
labeled ordering of vertices is determined by watching $k$ increase, from $1$
to $11$, as the algorithm progresses. \ We sweep across the tree horizontally
(in rows) rather than hierarchically. \ In particular, the command%
\[%
\begin{array}
[c]{l}%
\text{\texttt{k
$<$%
- k+(NROW(L1)!=0)+(NROW(L2)!=0)}}%
\end{array}
\]
increases $k$ by $2$ if both vertices are non-empty and by only $1$ if
otherwise.%
\begin{figure}[ptb]%
\centering
\includegraphics[
height=6.3362in,
width=3.9767in
]%
{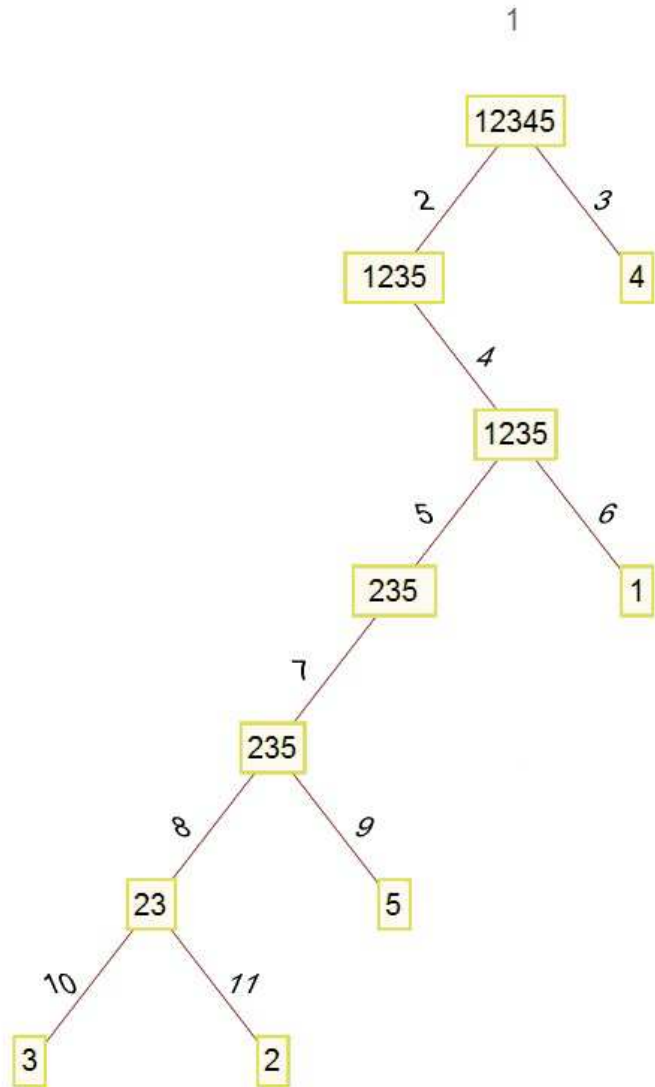}%
\caption{Weakly binary tree for leader election of size 11.}%
\end{figure}

The probability generating function for $Y_{n}$, given $n$, is \cite{Pr-tcs1}:%
\[%
\begin{array}
[c]{ccc}%
f_{n}(z)=-z^{2}2^{-n}+\left(  2z-z^{2}\right)  2^{-n}f_{n}(z)+z^{2}2^{-n}%
{\displaystyle\sum\limits_{k=0}^{n}}
\dbinom{n}{k}f_{k}(z), &  & n\geq2;
\end{array}
\]%
\[%
\begin{array}
[c]{ccc}%
f_{0}(z)=1, &  & f_{1}(z)=z.
\end{array}
\]
Note that $f_{n}(1)=1$ always. \ Differentiating with respect to $z$:%
\begin{align*}
f_{n}^{\prime}(z)  &  =-z2^{1-n}+\left(  1-z\right)  2^{1-n}f_{n}(z)+\left(
2z-z^{2}\right)  2^{-n}f_{n}^{\prime}(z)\\
&  +z2^{1-n}%
{\displaystyle\sum\limits_{k=0}^{n}}
\dbinom{n}{k}f_{k}(z)+z^{2}2^{-n}%
{\displaystyle\sum\limits_{k=0}^{n}}
\dbinom{n}{k}f_{k}^{\prime}(z)
\end{align*}
we have first moment%
\begin{align*}
\mathbb{E}(Y_{n})  &  =f_{n}^{\prime}(1)=-2^{1-n}+0+2^{-n}f_{n}^{\prime
}(1)+2+2^{-n}%
{\displaystyle\sum\limits_{k=0}^{n}}
\dbinom{n}{k}f_{k}^{\prime}(1)\\
&  =2-2^{1-n}+2^{-n}f_{n}^{\prime}(1)+2^{-n}%
{\displaystyle\sum\limits_{k=0}^{n}}
\dbinom{n}{k}f_{k}^{\prime}(1),
\end{align*}
that is,
\begin{align*}
g_{n}  &  =\left(  1-2^{1-n}\right)  +1+2^{-n}g_{n}+2^{-n}%
{\displaystyle\sum\limits_{k=0}^{n}}
\dbinom{n}{k}g_{k}\\
&  =1+\frac{1}{1-2^{1-n}}\left[  1+2^{-n}%
{\displaystyle\sum\limits_{k=0}^{n-1}}
\dbinom{n}{k}g_{k}\right]
\end{align*}
where $g_{k}=f_{k}^{\prime}(1)$ and $g_{0}=0$, $g_{1}=1$. \ Clearly $g_{2}=4$
and $g_{3}=29/6$. \ Differentiating again:%
\begin{align*}
f_{n}^{\prime\prime}(z)  &  =-2^{1-n}-2^{1-n}f_{n}(z)+(1-z)2^{2-n}%
f_{n}^{\prime}(z)+\left(  2z-z^{2}\right)  2^{-n}f_{n}^{\prime\prime}(z)\\
&  +2^{1-n}%
{\displaystyle\sum\limits_{k=0}^{n}}
\dbinom{n}{k}f_{k}(z)+z2^{2-n}%
{\displaystyle\sum\limits_{k=0}^{n}}
\dbinom{n}{k}f_{k}^{\prime}(z)+z^{2}2^{-n}%
{\displaystyle\sum\limits_{k=0}^{n}}
\dbinom{n}{k}f_{k}^{\prime\prime}(z)
\end{align*}
we have second factorial moment%
\begin{align*}
\mathbb{E}(Y_{n}(Y_{n}-1))  &  =f_{n}^{\prime\prime}(1)\\
&  =-2^{2-n}+0+2^{-n}f_{n}^{\prime\prime}(1)+2+2^{2-n}%
{\displaystyle\sum\limits_{k=0}^{n}}
\dbinom{n}{k}f_{k}^{\prime}(1)+2^{-n}%
{\displaystyle\sum\limits_{k=0}^{n}}
\dbinom{n}{k}f_{k}^{\prime\prime}(1),
\end{align*}
that is,
\begin{align*}
h_{n}  &  =2\left(  1-2^{1-n}\right)  +2^{-n}h_{n}+2^{2-n}%
{\displaystyle\sum\limits_{k=0}^{n}}
\dbinom{n}{k}g_{k}+2^{-n}%
{\displaystyle\sum\limits_{k=0}^{n}}
\dbinom{n}{k}h_{k}\\
&  =2+\frac{1}{1-2^{1-n}}\left[  2^{2-n}%
{\displaystyle\sum\limits_{k=0}^{n}}
\dbinom{n}{k}g_{k}+2^{-n}%
{\displaystyle\sum\limits_{k=0}^{n-1}}
\dbinom{n}{k}h_{k}\right]
\end{align*}
where $h_{k}=f_{k}^{\prime\prime}(1)$ and $h_{0}=h_{1}=0$. \ Clearly
$h_{2}=14$ and $h_{3}=200/9$. \ Finally, we have variance $\mathbb{V}(Y_{n})$
which is $2$ when $n=2$ and $133/36$ when $n=3$.

It can be proved that \cite{Pr-tcs1}%
\begin{align*}
\mathbb{E}(Y_{n})  &  =\frac{2\ln(n)}{\ln(2)}+\left(  2-\frac{\ln(\pi)-\gamma
}{\ln(2)}\right)  +\delta_{4}\left(  \frac{\ln(n)}{\ln(2)}\right)  +o(1)\\
&  \approx2\log_{2}(n)+(1.1812500478...)+\delta_{4}(\log_{2}(n))
\end{align*}
as $n\rightarrow\infty$, but a comparable asymptotic expression for
$\mathbb{V}(Y_{n})$ remains unknown.

We wonder about the cross-covariance of $H_{n}$ and $Y_{n}$, whose calculation
would require a bivariate generating function. \ 

\subsection{Draws}

\ Let us alter the rules so that a draw between two persons is allowed (if
precisely two persons are left, then they \textit{both} are declared leaders).
\ The third line of the two preceding R\ programs is simply replaced by%
\[%
\begin{array}
[c]{l}%
\text{\texttt{if(NROW(L)%
$<$%
=2) \{}}%
\end{array}
\]
and the initial conditions of the two preceding $\{g_{n},h_{n}\}$ recurrences
are also changed. \ For height, we now have $g_{0}=g_{1}=g_{2}=0$ (implying
$g_{3}=4/3)$ and $h_{0}=h_{1}=h_{2}=0$ (implying $h_{3}=8/9$); thus%
\begin{align*}
\mathbb{E}(\tilde{H}_{n})  &  =\frac{\ln(n)}{\ln(2)}+\left(  \frac{1}{2}%
-\frac{\pi^{2}}{12\ln(2)}\right)  +\delta_{5}\left(  \frac{\ln(n)}{\ln
(2)}\right)  +o(1)\\
&  \approx\log_{2}(n)-(0.6865691104...)+\delta_{5}(\log_{2}(n)).
\end{align*}
For size, we now have $g_{0}=0$, $g_{1}=g_{2}=1$ (implying $g_{3}=10/3)$ and
$h_{0}=h_{1}=h_{2}=0$ (implying $h_{3}=74/9$); thus%

\begin{align*}
\mathbb{E}(\tilde{Y}_{n})  &  =\frac{2\ln(n)}{\ln(2)}+\left(  2-\frac{\ln
(\pi)-\gamma+\frac{\pi^{2}}{8}}{\ln(2)}\right)  +\delta_{6}\left(  \frac
{\ln(n)}{\ln(2)}\right)  +o(1)\\
&  \approx2\log_{2}(n)-(0.5986036178...)+\delta_{6}(\log_{2}(n)).
\end{align*}
It would seem that the fraction $\pi^{2}/16$ given within the formula for
$\mathbb{E}(\tilde{Y}_{n})$ in \cite{Pr-tcs1} is incorrect and should be
$\pi^{2}/8$ instead.

\section{Coin Tosses}

Before moving on to more complicated examples, it may be worthwhile to
contemplate the most elementary recursive program imaginable:
\[%
\begin{array}
[c]{l}%
\text{\texttt{f
$<$%
- function(L,k)}}\\
\text{\texttt{\{}}\\
\text{\texttt{\ \ if(NROW(L)==0) \{}}\\
\text{\texttt{\ \ \ \ k}}\\
\text{\texttt{\ \ \} else \{}}\\
\text{\texttt{\ \ \ \ b
$<$%
- rbinom(NROW(L),1,0.5)}}\\
\text{\texttt{\ \ \ \ L1
$<$%
- L[(1:NROW(L))*(1-b)]}}\\
\text{\texttt{\ \ \ \ k
$<$%
- f(L1,k
$<$%
- k+1)}}\\
\text{\texttt{\ \ \}}}\\
\text{\texttt{\ \ k}}\\
\text{\texttt{\}}}%
\end{array}
\]
mentioned at the end of \cite{Fi1-tcs1}, where $k$ is initially $0$. \ Note
that the third line here is%
\[%
\begin{array}
[c]{l}%
\text{\texttt{if(NROW(L)%
$<$%
=0) \{}}%
\end{array}
\]
unlike what we have seen before. \ People iteratively toss all coins which
show tails until only heads are visible. Omitting details, it follows that the
procedural height has generating function \
\[%
\begin{array}
[c]{ccc}%
f_{n}(z)=z2^{-n}%
{\displaystyle\sum\limits_{k=0}^{n}}
\dbinom{n}{k}f_{k}(z), &  & n\geq1;
\end{array}
\]%
\[
f_{0}(z)=1
\]
therefore%
\[
g_{n}=1+2^{-n}%
{\displaystyle\sum\limits_{k=0}^{n}}
\dbinom{n}{k}g_{k}=\frac{1}{1-2^{-n}}\left[  1+2^{-n}%
{\displaystyle\sum\limits_{k=0}^{n-1}}
\dbinom{n}{k}g_{k}\right]  ,
\]%
\[
h_{n}=-2+2g_{n}+2^{-n}%
{\displaystyle\sum\limits_{k=0}^{n}}
\dbinom{n}{k}h_{k}=\frac{1}{1-2^{-n}}\left[  -2+2g_{n}+2^{-n}%
{\displaystyle\sum\limits_{k=0}^{n-1}}
\dbinom{n}{k}h_{k}\right]
\]
where $g_{0}=0$ (implying $g_{1}=2$, $g_{2}=8/3$, $g_{3}=22/7)$ and $h_{0}=0$
(implying $h_{1}=4$, $h_{2}=64/9$, $h_{3}=1420/147$). \ From this
\cite{SR-tcs1}%
\begin{align*}
\mathbb{E}(\hat{H}_{n})  &  =\frac{\ln(n)}{\ln(2)}+\left(  \frac{1}{2}%
+\frac{\gamma}{\ln(2)}\right)  +\delta_{7}\left(  \frac{\ln(n)}{\ln
(2)}\right)  +o(1)\\
&  \approx\log_{2}(n)+(1.3327461772...)+\delta_{7}(\log_{2}(n)),
\end{align*}%
\begin{align*}
\mathbb{V}(\hat{H}_{n})  &  =\left(  \frac{1}{12}+\frac{\pi^{2}}{6\ln(2)^{2}%
}\right)  +\delta_{8}\left(  \frac{\ln(n)}{\ln(2)}\right)  +o(1)\\
&  \approx(3.5070480758...)+\,\delta_{8}\left(  \log_{2}(n)\right)
\end{align*}
as $n\rightarrow\infty$.

\section{Finding the Maximum}

Consider the R\ program:%
\[%
\begin{array}
[c]{l}%
\text{\texttt{f
$<$%
- function(L,k,r)}}\\
\text{\texttt{\{}}\\
\text{\texttt{\ \ if(NROW(L)==0) \{}}\\
\text{\texttt{\ \ \ \ c(k,r)}}\\
\text{\texttt{\ \ \} else if(NROW(L)==1) \{}}\\
\text{\texttt{\ \ \ \ c(k,r
$<$%
- max(L,r))}}\\
\text{\texttt{\ \ \} else \{}}\\
\text{\texttt{\ \ \ \ b
$<$%
- rbinom(NROW(L),1,0.5)}}\\
\text{\texttt{\ \ \ \ L0
$<$%
- L[(1:NROW(L))*b]}}\\
\text{\texttt{\ \ \ \ L1
$<$%
- L[(1:NROW(L))*(1-b)]}}\\
\text{\texttt{\ \ \ \ V
$<$%
- f(L1,k
$<$%
- k+1,r)}}\\
\text{\texttt{\ \ \ \ k
$<$%
- V[1]; r
$<$%
- V[2]}}\\
\text{\texttt{\ \ \ \ L2
$<$%
- L0[L0%
$>$%
r]}}\\
\text{\texttt{\ \ \ \ V
$<$%
- f(L2,k
$<$%
- k+1,r)}}\\
\text{\texttt{\ \ \ \ k
$<$%
- V[1]; r
$<$%
- V[2]}}\\
\text{\texttt{\ \ \}}}\\
\text{\texttt{\ \ c(k,r)}}\\
\text{\texttt{\}}}%
\end{array}
\]
where the integer $k$ is initially $0$ and the list $L$ is initially
$\{1,2,\ldots,n\}$. \ We find a comparison of a conflict resolution tree in
Figure 4 and its corresponding maximum-finding tree in Figure 5 to be helpful.
\ Note that the vertex count for the latter is only $14$, which is less than
the vertex count of $16$ determined for the former. \ The discrepancy arises
because, at the instant the $5^{\text{th}}$ vertex $3$ becomes\ the leader and
$r$ consequently increases from $0$ to $3$,\ both the $6^{\text{th}}$ vertex
$1$ becomes empty and the $8^{\text{th}}$ \& $9^{\text{th}}$ vertices $2456$
are reduced to $456$. \ It follows that the $10^{\text{th}}$ vertex $24$ is
reduced to $4$, which cannot have any descendants, eliminating two twigs.
\ Also, $r$ increases from $3$ to $4$ and, subsequently, from $4$ to $6$;
while the penultimate vertex becomes empty, this does not further alter the
count. \ We write $X_{6}=14$.%
\begin{figure}[ptb]%
\centering
\includegraphics[
height=6.3362in,
width=6.8692in
]%
{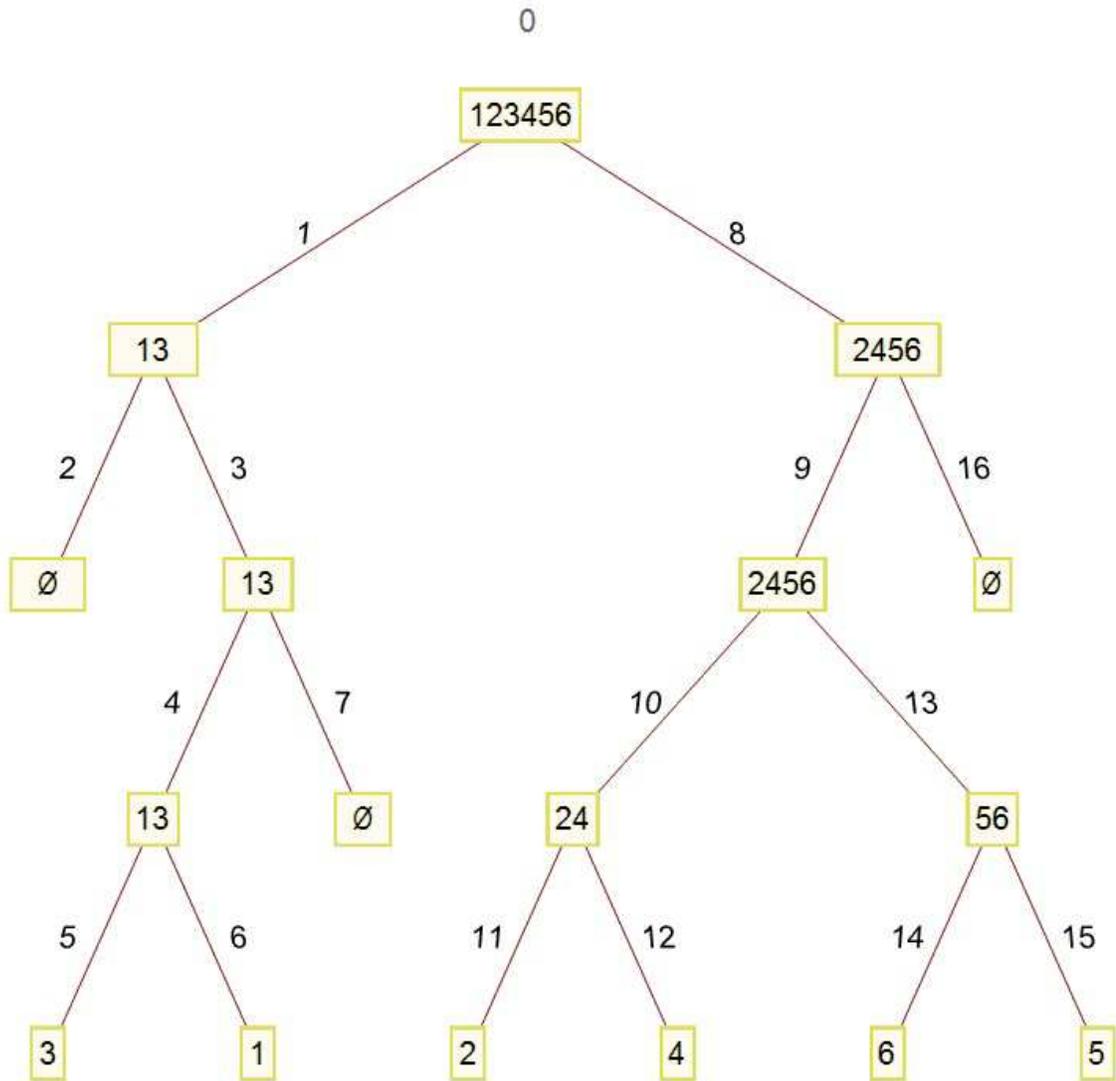}%
\caption{Conflict resolution tree (to be compared with next figure).}%
\end{figure}
\begin{figure}[ptb]%
\centering
\includegraphics[
height=6.3387in,
width=6.1619in
]%
{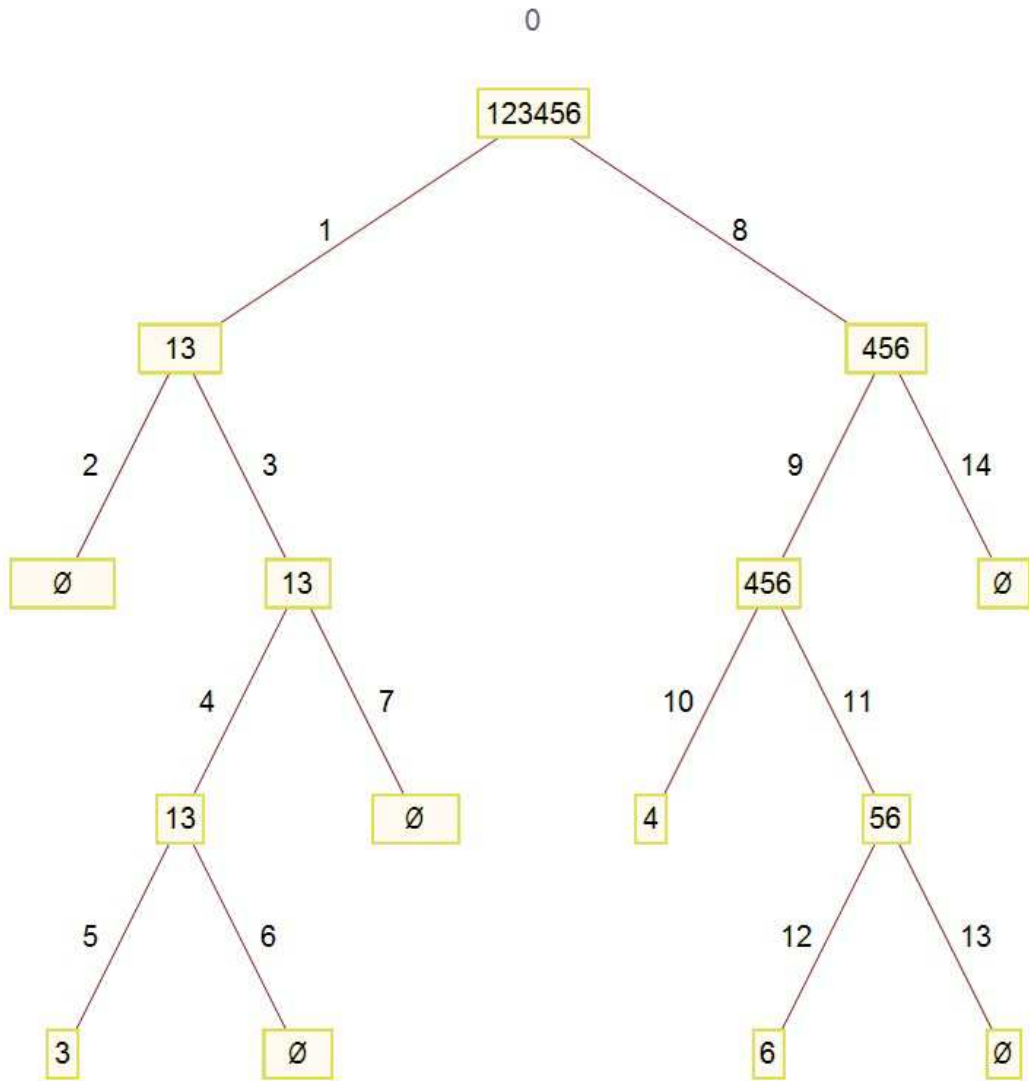}%
\caption{Maximum-finding tree (to be compared with previous figure).}%
\end{figure}

\newpage

The probability generating function for $X_{n}$, given $n$, is \cite{SY0-tcs1,
SY1-tcs1, GP1-tcs1, CH-tcs1}:%
\[%
\begin{array}
[c]{ccc}%
f_{n}(z)=z^{2}2^{-n}f_{n}(z)+z2^{-n}%
{\displaystyle\sum\limits_{k=1}^{n}}
f_{n-k}(z)%
{\displaystyle\sum\limits_{j=1}^{k}}
\dbinom{k-1}{j-1}f_{j}(z), &  & n\geq2;
\end{array}
\]%
\[
f_{0}(z)=f_{1}(z)=z.
\]
Note that $f_{n}(1)=1$ always. \ When differentiating with respect to $z$:%
\begin{align*}
f_{n}^{\prime}(z)  &  =z2^{1-n}f_{n}(z)+z^{2}2^{-n}f_{n}^{\prime}(z)+2^{-n}%
{\displaystyle\sum\limits_{k=1}^{n}}
f_{n-k}(z)%
{\displaystyle\sum\limits_{j=1}^{k}}
\dbinom{k-1}{j-1}f_{j}(z)\\
&  +z2^{-n}%
{\displaystyle\sum\limits_{k=1}^{n}}
f_{n-k}^{\prime}(z)%
{\displaystyle\sum\limits_{j=1}^{k}}
\dbinom{k-1}{j-1}f_{j}(z)+z2^{-n}%
{\displaystyle\sum\limits_{j=1}^{n}}
f_{j}^{\prime}(z)%
{\displaystyle\sum\limits_{k=j}^{n}}
\dbinom{k-1}{j-1}f_{n-k}(z)
\end{align*}
we make use of the identity%
\[%
{\displaystyle\sum\limits_{k=1}^{n}}
{\displaystyle\sum\limits_{j=1}^{k}}
a_{kj}=%
{\displaystyle\sum\limits_{j=1}^{n}}
{\displaystyle\sum\limits_{k=j}^{n}}
a_{kj}.
\]
From this follows the first moment:%
\begin{align*}
\mathbb{E}(X_{n})  &  =f_{n}^{\prime}(1)=2^{1-n}+2^{-n}f_{n}^{\prime
}(1)+2^{-n}\left(  -1+2^{n}\right)  +2^{-n}%
{\displaystyle\sum\limits_{k=1}^{n}}
2^{-1+k}f_{n-k}^{\prime}(1)+2^{-n}%
{\displaystyle\sum\limits_{j=1}^{n}}
\dbinom{n}{j}f_{j}^{\prime}(1)\\
&  =1-2^{-n}+2^{1-n}+2^{-n}f_{n}^{\prime}(1)+%
{\displaystyle\sum\limits_{k=1}^{n}}
2^{-1-n+k}f_{n-k}^{\prime}(1)+2^{-n}%
{\displaystyle\sum\limits_{j=0}^{n-1}}
\dbinom{n}{j}f_{j}^{\prime}(1)-2^{-n}+2^{-n}f_{n}^{\prime}(1)\\
&  =1+2^{1-n}f_{n}^{\prime}(1)+%
{\displaystyle\sum\limits_{i=0}^{n-1}}
2^{-i-1}f_{i}^{\prime}(1)+2^{-n}%
{\displaystyle\sum\limits_{j=0}^{n-1}}
\dbinom{n}{j}f_{j}^{\prime}(1)
\end{align*}
setting $i=n-k$, hence $i+1=1+n-k$ and $0\leq i\leq n-1$ when $1\leq k\leq n$.
\ That is,%
\begin{align*}
g_{n}  &  =1+2^{1-n}g_{n}+%
{\displaystyle\sum\limits_{i=0}^{n-1}}
\left[  2^{-n}\dbinom{n}{i}+2^{-i-1}\right]  g_{i}\\
&  =\frac{1}{1-2^{1-n}}\left\{  1+%
{\displaystyle\sum\limits_{i=0}^{n-1}}
\left[  2^{-n}\dbinom{n}{i}+2^{-i-1}\right]  g_{i}\right\}
\end{align*}
where $g_{i}=f_{i}^{\prime}(1)$ and $g_{0}=g_{1}=1$. \ Clearly $g_{2}=5$ and
$g_{3}=19/3$. \ Differentiating again:
\begin{align*}
f_{n}^{\prime\prime}(z)  &  =2^{1-n}f_{n}(z)+z2^{2-n}f_{n}^{\prime}%
(z)+z^{2}2^{-n}f_{n}^{\prime\prime}(z)\\
&  +2^{1-n}%
{\displaystyle\sum\limits_{k=1}^{n}}
f_{n-k}^{\prime}(z)%
{\displaystyle\sum\limits_{j=1}^{k}}
\dbinom{k-1}{j-1}f_{j}(z)+2^{1-n}%
{\displaystyle\sum\limits_{j=1}^{n}}
f_{j}^{\prime}(z)%
{\displaystyle\sum\limits_{k=j}^{n}}
\dbinom{k-1}{j-1}f_{n-k}(z)\\
&  +z2^{-n}%
{\displaystyle\sum\limits_{k=1}^{n}}
f_{n-k}^{\prime\prime}(z)%
{\displaystyle\sum\limits_{j=1}^{k}}
\dbinom{k-1}{j-1}f_{j}(z)+z2^{-n}%
{\displaystyle\sum\limits_{j=1}^{n}}
f_{j}^{\prime\prime}(z)%
{\displaystyle\sum\limits_{k=j}^{n}}
\dbinom{k-1}{j-1}f_{n-k}(z)\\
&  +z2^{1-n}%
{\displaystyle\sum\limits_{k=1}^{n}}
f_{n-k}^{\prime}(z)%
{\displaystyle\sum\limits_{j=1}^{k}}
\dbinom{k-1}{j-1}f_{j}^{\prime}(z)
\end{align*}
and employing%
\begin{align*}
2^{-n}%
{\displaystyle\sum\limits_{k=1}^{n}}
2^{-1+k}f_{n-k}^{\prime}(1)+2^{-n}%
{\displaystyle\sum\limits_{j=1}^{n}}
\dbinom{n}{j}f_{j}^{\prime}(1)  &  =f_{n}^{\prime}(1)-\left[  1-2^{-n}%
+2^{1-n}+2^{-n}f_{n}^{\prime}(1)\right] \\
&  =-1-2^{-n}+\left(  1-2^{-n}\right)  f_{n}^{\prime}(1),
\end{align*}
we have second factorial moment%
\begin{align*}
\mathbb{E}(X_{n}(X_{n}-1))  &  =f_{n}^{\prime\prime}(1)\\
&  =2^{1-n}+2^{2-n}f_{n}^{\prime}(1)+2^{-n}f_{n}^{\prime\prime}(1)+2\left[
-1-2^{-n}+\left(  1-2^{-n}\right)  f_{n}^{\prime}(1)\right] \\
&  +%
{\displaystyle\sum\limits_{i=0}^{n-1}}
2^{-i-1}f_{i}^{\prime\prime}(1)+%
{\displaystyle\sum\limits_{i=0}^{n-1}}
2^{-n}\dbinom{n}{i}f_{i}^{\prime\prime}(1)+2^{-n}f_{n}^{\prime\prime
}(1)+2^{1-n}%
{\displaystyle\sum\limits_{k=1}^{n}}
f_{n-k}^{\prime}(1)%
{\displaystyle\sum\limits_{j=1}^{k}}
\dbinom{k-1}{j-1}f_{j}^{\prime}(1)
\end{align*}
that is,%
\begin{align*}
h_{n}  &  =-2+2\left(  1+2^{-n}\right)  g_{n}+2^{1-n}h_{n}+%
{\displaystyle\sum\limits_{i=0}^{n-1}}
\left[  2^{-n}\dbinom{n}{i}+2^{-i-1}\right]  h_{i}+2^{1-n}%
{\displaystyle\sum\limits_{k=1}^{n}}
g_{n-k}%
{\displaystyle\sum\limits_{j=1}^{k}}
\dbinom{k-1}{j-1}g_{j}\\
&  =\frac{1}{1-2^{1-n}}\left\{  -2+2\left(  1+2^{-n}\right)  g_{n}+%
{\displaystyle\sum\limits_{i=0}^{n-1}}
\left[  2^{-n}\dbinom{n}{i}+2^{-i-1}\right]  h_{i}+2^{1-n}%
{\displaystyle\sum\limits_{k=1}^{n}}
{\displaystyle\sum\limits_{j=1}^{k}}
\dbinom{k-1}{j-1}g_{j}g_{n-k}\right\}
\end{align*}
where $h_{k}=f_{k}^{\prime\prime}(1)$ and $h_{0}=h_{1}=0$. \ Clearly
$h_{2}=28$ and $h_{3}=400/9$. \ Finally, we have variance $\mathbb{V}(X_{n})$
which is $8$ when $n=2$ and $32/3$ when $n=3$.

It can be proved that \cite{GP1-tcs1, CH-tcs1}
\begin{align*}
\mathbb{E}(X_{n})  &  =\frac{\pi^{2}}{3\ln(2)}\ln(n)+\delta_{9}\left(
\frac{\ln(n)}{\ln(2)}\right)  +o(1)\\
&  \approx(4.7462764416...)\ln(n)+\delta_{9}(\log_{2}(n)),
\end{align*}%
\begin{align*}
\mathbb{V}(X_{n})  &  =\frac{1}{\ln(2)}\left(  \pi^{2}-2-\frac{\pi^{4}}{9}+%
{\displaystyle\sum\limits_{i=0}^{\infty}}
2^{-i-1}h_{i}\right)  \ln(n)+O(1)\\
&  \approx(11.7013270183...)\ln(n)+\,O(1)
\end{align*}
as $n\rightarrow\infty$. \ Although the leading coefficient for $\mathbb{V}%
(X_{n})$ is recursive (involving $h_{i}$), the series converges rapidly enough
for numerical purposes.

Shiau \&\ Yang \cite{SY2-tcs1} mentioned a slight revision of the preceding
algorithm, but details are unclear. \ To discuss this, we write $Y_{n}$ for
the vertex count rather than $X_{n}$. \ In particular, they gave
$\mathbb{E}(Y_{2})=9/2$. \ We hypothesize that, in the maximization R\ code
given earlier, replacing the line
\[%
\begin{array}
[c]{l}%
\text{\texttt{V
$<$%
- f(L2,k
$<$%
- k+1,r)}}%
\end{array}
\]
by the two lines:%
\[%
\begin{array}
[c]{l}%
\text{\texttt{if(NROW(L1)%
$>$%
=1) k
$<$%
- k+1}}\\
\text{\texttt{V
$<$%
- f(L2,k,r)}}%
\end{array}
\]
might duplicate the revised procedure. This has the effect of eliminating the
$2^{\text{nd}}$ vertex $\emptyset$ in Figure 5 (because the $3^{\text{rd}}$
vertex $13$ will be assigned the same label $2$). \ It makes sense to remove
left-empty vertices from consideration for efficiency (as would removing
right-empty vertices, or both). Simulation suggests that, indeed, the average
vertex count is $9/2$ when $n=2$. \ We have not further pursued this issue.

Chen \& Hwang \cite{CH-tcs1} analyzed a different algorithm, based on leader
elections, that starts afresh with every iteration. \ It is not surprising
that vertex counts are $O\left(  \ln(n)^{2}\right)  $ for the mean and
$O\left(  \ln(n)^{3}\right)  $ for the variance, notably slower than the
conflict resolution-based approach.

\section{Sorting a List}

Consider the R\ programs:%
\[%
\begin{array}
[c]{l}%
\text{\texttt{phi
$<$%
- function(L,k,s)}}\\
\text{\texttt{\{}}\\
\text{\texttt{\ \ if(NROW(L)==0) \{}}\\
\text{\texttt{\ \ \ \ c(k,s)}}\\
\text{\texttt{\ \ \} else if(NROW(L)==1) \{}}\\
\text{\texttt{\ \ \ \ c(k,s
$<$%
- L[1])}}\\
\text{\texttt{\ \ \} else \{}}\\
\text{\texttt{\ \ \ \ b
$<$%
- rbinom(NROW(L),1,0.5)}}\\
\text{\texttt{\ \ \ \ L1
$<$%
- L[(1:NROW(L))*(1-b)]}}\\
\text{\texttt{\ \ \ \ V
$<$%
- phi(L1,k
$<$%
- k+1,s)}}\\
\text{\texttt{\ \ \ \ k
$<$%
- V[1]; s
$<$%
- V[2]}}\\
\text{\texttt{\ \ \ \ if(NROW(L1)==0) \{}}\\
\text{\texttt{\ \ \ \ \ \ L2
$<$%
- L[(1:NROW(L))*b]}}\\
\text{\texttt{\ \ \ \ \ \ V
$<$%
- phi(L2,k,s)}}\\
\text{\texttt{\ \ \ \ \ \ k
$<$%
- V[1]; s
$<$%
- V[2]}}\\
\text{\texttt{\ \ \ \ \}}}\\
\text{\texttt{\ \ \}}}\\
\text{\texttt{\ \ c(k,s)}}\\
\text{\texttt{\}}}%
\end{array}
\]
and%
\[%
\begin{array}
[c]{l}%
\text{\texttt{psi
$<$%
- function(L)}}\\
\text{\texttt{\{}}\\
\text{\texttt{\ \ if(NROW(L)%
$<$%
=1) \{}}\\
\text{\texttt{\ \ \ \ k
$<$%
- 1}}\\
\text{\texttt{\ \ \} else \{}}\\
\text{\texttt{\ \ \ \ V
$<$%
- phi(L,0,0)}}\\
\text{\texttt{\ \ \ \ k
$<$%
- 1+V[1]; s
$<$%
- V[2]}}\\
\text{\texttt{\ \ \ \ k
$<$%
- k+psi(L[L%
$<$%
s])+psi(L[L%
$>$%
s])}}\\
\text{\texttt{\ \ \}}}\\
\text{\texttt{\ \ k}}\\
\text{\texttt{\}}}%
\end{array}
\]
where the list $L$ is initially $\{1,2,\ldots,n\}$. \ The function $\varphi$
accomplishes the same task as $f$ does in Section 2.1, but additionally
monitors the leader $s$, whose value acts as a pivot for the recursive
splitting of $L$ within the function $\psi$. \ \ Define $K_{n}$ to be the
\textquotedblleft height\textquotedblright\ of the sorting algorithm captured
by $\psi$, that is, the sum (over all elections) of required time lengths plus
one. \ 

The probability generating function for $K_{n}$, given $n$, is%
\[%
\begin{array}
[c]{ccc}%
\psi_{n}(z)=\dfrac{zf_{n}(z)}{n}%
{\displaystyle\sum\limits_{k=0}^{n-1}}
\psi_{k}(z)\psi_{n-k-1}(z), &  & n\geq2;
\end{array}
\]%
\[
\psi_{0}(z)=\psi_{1}(z)=z
\]
where $f_{n}$ is exactly as in Section 2.1. \ Note that $\psi_{n}(1)=1$
always. \ Differentiating with respect to $z$:%
\[
\psi_{n}^{\prime}(z)=\dfrac{f_{n}(z)+zf_{n}^{\prime}(z)}{n}%
{\displaystyle\sum\limits_{k=0}^{n-1}}
\psi_{k}(z)\psi_{n-k-1}(z)+\dfrac{zf_{n}(z)}{n}%
{\displaystyle\sum\limits_{k=0}^{n-1}}
\left[  \psi_{k}^{\prime}(z)\psi_{n-k-1}(z)+\psi_{k}(z)\psi_{n-k-1}^{\prime
}(z)\right]
\]
we have first moment%
\begin{align*}
\mathbb{E}(K_{n})  &  =\psi_{n}^{\prime}(1)=\dfrac{1+f_{n}^{\prime}(1)}{n}%
{\displaystyle\sum\limits_{k=0}^{n-1}}
1+\dfrac{1}{n}%
{\displaystyle\sum\limits_{k=0}^{n-1}}
\left[  \psi_{k}^{\prime}(1)+\psi_{n-k-1}^{\prime}(1)\right] \\
&  =1+f_{n}^{\prime}(1)+\dfrac{2}{n}%
{\displaystyle\sum\limits_{k=0}^{n-1}}
\psi_{k}^{\prime}(1),
\end{align*}
that is \cite{SY3-tcs1, GP2-tcs1},
\[
\xi_{n}=1+g_{n}+\dfrac{2}{n}%
{\displaystyle\sum\limits_{k=0}^{n-1}}
\xi_{k}%
\]
where $g_{k}=f_{k}^{\prime}(1)$, $\xi_{k}=\psi_{k}^{\prime}(1)$ and $\xi
_{0}=\xi_{1}=1$. \ Clearly $\xi_{2}=5$ and $\xi_{3}=8$. \ Differentiating
again:%
\begin{align*}
\psi_{n}^{\prime\prime}(z)  &  =\dfrac{2f_{n}^{\prime}(z)+zf_{n}^{\prime
\prime}(z)}{n}%
{\displaystyle\sum\limits_{k=0}^{n}}
\psi_{k}(z)\psi_{n-k-1}(z)\\
&  +\dfrac{2f_{n}(z)+2zf_{n}^{\prime}(z)}{n}%
{\displaystyle\sum\limits_{k=0}^{n-1}}
\left[  \psi_{k}^{\prime}(z)\psi_{n-k-1}(z)+\psi_{k}(z)\psi_{n-k-1}^{\prime
}(z)\right] \\
&  +\dfrac{zf_{n}(z)}{n}%
{\displaystyle\sum\limits_{k=0}^{n-1}}
\left[  \psi_{k}^{\prime\prime}(z)\psi_{n-k-1}(z)+2\psi_{k}^{\prime}%
(z)\psi_{n-k-1}^{\prime}(z)+\psi_{k}(z)\psi_{n-k-1}^{\prime\prime}(z)\right]
\end{align*}
we have second factorial moment%
\begin{align*}
\mathbb{E}(K_{n}(K_{n}-1))  &  =\psi_{n}^{\prime\prime}(1)\\
&  =\dfrac{2f_{n}^{\prime}(1)+f_{n}^{\prime\prime}(1)}{n}%
{\displaystyle\sum\limits_{k=0}^{n}}
1+\dfrac{2+2f_{n}^{\prime}(1)}{n}%
{\displaystyle\sum\limits_{k=0}^{n-1}}
\left[  \psi_{k}^{\prime}(1)+\psi_{n-k-1}^{\prime}(1)\right] \\
&  +\dfrac{1}{n}%
{\displaystyle\sum\limits_{k=0}^{n-1}}
\left[  \psi_{k}^{\prime\prime}(1)+2\psi_{k}^{\prime}(1)\psi_{n-k-1}^{\prime
}(1)+\psi_{n-k-1}^{\prime\prime}(1)\right] \\
&  =2f_{n}^{\prime}(1)+f_{n}^{\prime\prime}(1)+\dfrac{4+4f_{n}^{\prime}(1)}{n}%
{\displaystyle\sum\limits_{k=0}^{n-1}}
\psi_{k}^{\prime}(1)+\dfrac{2}{n}%
{\displaystyle\sum\limits_{k=0}^{n-1}}
\left[  \psi_{k}^{\prime\prime}(1)+\psi_{k}^{\prime}(1)\psi_{n-k-1}^{\prime
}(1)\right]
\end{align*}
that is,
\[
\eta_{n}=2g_{n}+h_{n}-2\left(  1+g_{n}\right)  \left(  1+g_{n}-\xi_{n}\right)
+\dfrac{2}{n}%
{\displaystyle\sum\limits_{k=0}^{n-1}}
\left(  \eta_{k}+\xi_{k}\xi_{n-k-1}\right)
\]
where $h_{k}=f_{k}^{\prime\prime}(1)$, $\eta_{k}=\psi_{k}^{\prime\prime}(1)$
and $\eta_{0}=\eta_{1}=0$. \ As far as is known, this recurrence has not
previously appeared in the literature. \ Clearly $\eta_{2}=22$ and $\eta
_{3}=190/3$. \ Finally, we have variance $\mathbb{V}(K_{n})$ which is $2$ when
$n=2$ and $22/3$ when $n=3$. \ It can be proved that \cite{GP2-tcs1}%
\begin{align*}
\frac{\mathbb{E}(K_{n})}{n}  &  =\frac{8}{3}+2%
{\displaystyle\sum\limits_{\ell=3}^{\infty}}
\frac{g_{\ell}-g_{\ell-1}}{\ell+1}+o(1)\\
&  \approx3.5455178132...\approx8/3+2(0.4394255733...)
\end{align*}
as $n\rightarrow\infty$, but a comparable limit for $\mathbb{V}(K_{n})$
remains unknown.

Grabner \&\ Prodinger \cite{GP2-tcs1} examined a different sorting algorithm
\cite{SY2-tcs1, SY4-tcs1}, based on the preceding maximization strategy, that
is somewhat more involved yet does not perform quite as well:%
\begin{align*}
\frac{\mathbb{E}(\tilde{K}_{n})}{n} &  =\frac{13}{6}+%
{\displaystyle\sum\limits_{\ell=3}^{\infty}}
\frac{g_{\ell}-g_{\ell-1}}{\ell+1}+o(1)\\
&  \approx3.6798261095...\approx13/6+2(0.7565797214...)
\end{align*}
as $n\rightarrow\infty$, where $g_{n}$ is exactly as in Section 4. \ A third,
naive algorithm was also examined that is $O\left(  n\ln(n)\right)  $; an
interesting constant%
\begin{align*}
2+%
{\displaystyle\sum\limits_{\ell=1}^{\infty}}
\left(  1-2^{\ell}%
{\displaystyle\sum\limits_{m=2^{\ell}}^{\infty}}
\frac{1}{m^{2}}\right)   &  =1.4463764113...\\
&  =-\frac{3}{4}+\frac{1}{\ln(2)}+0.7536813704...
\end{align*}
arises and plays a role in the fifth term of the performance asymptotics. \ We
observe a passing resemblance between this and another constant \cite{FN-tcs1,
GP3-tcs1, CFV-tcs1}%
\[
2%
{\displaystyle\sum\limits_{\ell=0}^{\infty}}
\left(  1+\frac{1}{2^{\ell}}%
{\displaystyle\sum\limits_{m=1}^{2^{\ell}}}
\ln\left(  \frac{m}{2^{\ell}}\right)  \right)  =5.2793782410...
\]
that occurs elsewhere; a concise expression for an associated quantity
$8.2073088638...$ remains open.

\section{Closing Words}

Can algorithmic optimality can be demonstrated for any of the topics discussed
in this paper? \ We suspect that such a proof might be more feasible for
conflict resolution than for maximum finding or list sorting. \ A related
topic has to do with \textbf{initialization}, that is, the assignment of a
unique identifier to each of the $n$ individuals \cite{NO-tcs1, SY5-tcs1,
IW-tcs1}, which we omit for reasons of time and space.

Instead of the pool of $n$ persons being fixed, let us imagine it as variable.
\ Assume that people appear gradually according to a Poisson process in
continuous time with arrival rate $\lambda$. Upon appearance, they attempt to
broadcast their data at the start of the next available integer time slot,
i.e., access for newcomers is free. \ What is the largest value $\lambda$ for
which all ensuing conflicts are resolved in finite time almost surely? \ The
critical point $\lambda_{c}$ is called the \textbf{maximum stable throughput}
of the algorithm. \ It is known to satisfy a transcendental equation
\cite{MF-tcs1, FHJ-tcs1, PV-tcs1}%
\[
\frac{q(1-\lambda)-1}{\lambda q^{2}}\exp\left[  \frac{q\lambda}{q-1}\right]  =%
{\displaystyle\sum\limits_{k=1}^{\infty}}
\frac{k}{k+1}\left(  k\frac{1-q^{-1}}{1-q^{-k}}-\frac{1}{q}\right)  \left(
\frac{q}{q-1}\right)  ^{k}\frac{\lambda^{k}}{k!}%
\]
with $q=2$ and $\lambda_{c}=0.3601770279...$. \ Also, if the coins are 3-sided
rather than 2-sided, then the equation is applicable with $q=3$ and
$\lambda_{c}=0.4015993701...$; if instead they are 4-sided, then $q=4$ applies
and $\lambda_{c}=0.3992228263...$. \ Hence ternary trees are better than both
binary trees and quaternary trees with regard to maximum throughput. \ The
same is true when access is blocked for newcomers:\ $\hat{\lambda}_{c}%
=\ln(q)/q$ is largest when the integer $q$ is $3$. \ A wealth of numerical
information is given in \cite{MF-tcs1}, but we must stop here.

\section{Acknowledgements}

Helpful correspondence with Hsien-Kuei Hwang, Shyue-Horng Shiau, Helmut
Prodinger and Svante Janson is greatly appreciated. \

\end{document}